\documentclass[11pt,a4paper]{article}
\usepackage[utf8]{inputenc}
\usepackage[british]{babel}
\usepackage{amsmath,amssymb,amsfonts,bm}
\usepackage[margin=2.4cm]{geometry}
\usepackage{graphicx,float}
\usepackage[font=small,labelfont=bf]{caption}

\usepackage[superscript]{cite}
\usepackage{xcolor}
\definecolor{linkcol}{RGB}{30,70,120}
\usepackage[colorlinks=true,allcolors=linkcol]{hyperref}
\newcommand{\K}{\mathbf{K}}
\newcommand{\R}{\mathbf{R}}
\newcommand{\Mc}{\mathcal{M}}
\newcommand{\Lc}{\mathcal{L}}
\newcommand{\avg}[1]{\left\langle #1 \right\rangle}
\newcommand{\Keff}{K a\cos\tilde\beta}

\title{\textbf{The Ising lattice gas and the surface diffusion\\ of CO on Cu(001) at finite coverage}}
\author{S. Miret-Art\'es\\[4pt]
\small Instituto de F\'isica Fundamental, CSIC, Serrano 123, 28006 Madrid, Spain}
\date{}

\begin{document}
\maketitle

\begin{abstract}
\noindent
The memory equation of the coherent intermediate scattering function (ISF),
derived previously for lattice gases, is applied to the Ising lattice gas under
Kawasaki dynamics. The ISF follows the decay of a density wave of adparticles, a
superposition of exponential relaxation modes: the decay starts at a rate fixed
by a sum rule and is then slowed by a memory. The static structure factor and
the effective hop rate are obtained in closed form, exactly along a
one-dimensional channel; with the sum rule they give a single exponential, a
lower bound on the ISF.
Exact relaxation spectra show that this single exponential, built from
equilibrium quantities alone, reproduces the correlation time only at weak
coupling; only there can a measured dephasing rate be read through the statics. For strong attraction the slow
dissolution of clusters lengthens the correlation time up to ninefold, whereas a
repulsive layer relaxes nearly as a single mode at its structure peak. The dephasing rate,
fitted over a time window where it is nearly constant, approaches the slowest
relaxation rate, not the initial one; the sum rule reaches the data only through
the area beneath the normalized decay, which defines the correlation time. For CO/Cu(001), the lattice gas reproduces the dephasing rates
measured along two crystal directions and requires a nearest-neighbour repulsion
of at least $35$--$50$~meV, whatever the jump rates; the measured rate is then
close to the inverse correlation time, and the memory is predicted to grow with
coverage.
\end{abstract}

\section{Introduction}
\label{sec:intro}

Helium spin-echo (HeSE) probes surface diffusion in the time domain with
atomic-scale resolution\cite{Jardine2009}. It returns the coherent ISF, $I(\K,t)$, at the momentum transfer $\K$ parallel to
the surface. The ISF is a characteristic function (CF) in the strict sense of probability
theory. The CF framework built on that property has organized the description of
single-adsorbate diffusion\cite{TorresMiyares2026a,TorresMiyares2026b,TorresMiyares2026c},
from the compound-Poisson structure of Markovian jumps\cite{Montroll1965} to the
hierarchy of linear-response functions\cite{TorresMiyares2026c}.
What the framework has lacked is the lateral interaction between adparticles.
The so-called interacting-single-adsorbate model reduced it to a collisional
friction\cite{MartinezCasado2007}, which retains neither its sign nor the
correlations it induces\cite{Ellis2001,Alexandrowicz2006}. The lattice-gas
tradition\cite{Reed1981,Gomer1990,Uebing1991,AlaNissila2002,Gortel2004,Tarasenko2018}
has treated interacting diffusion mainly through diffusion coefficients, not
through the time- and momentum-resolved observable.

In a previous work\cite{PaperI}, the lateral interaction was treated for a whole
class of lattice gases, the class meant throughout the present work. The
adsorption sites form a Bravais lattice, and each site $\R$ is empty or holds
one adparticle, as recorded by the occupation number $n_{\R}=0$ or $1$. The set
of all occupation numbers is the configuration of the adlayer, and the mean
occupation per site is the coverage $\theta$.
A member of the class has four properties: (i) the number of adparticles is
conserved; (ii) an adparticle jumps to an empty site along one of a fixed set
of jump vectors; (iii) the jump rates, which may depend on the occupations
around the jump, obey detailed balance with respect to the equilibrium
distribution of configurations; and (iv) the dynamics is invariant under the
lattice translations and inversion. The lateral interaction is otherwise
unrestricted.
For every member of the class, the ISF obeys an exact memory equation. Its decay
sets out at a local or initial rate, fixed by an exact sum rule in terms of equilibrium
quantities. It is slowed afterwards by a memory. The
decay of a density wave is a superposition of relaxation modes, each decaying exponentially with
its own relaxation time, and all entering with non-negative weights. The
relaxation spectrum is the distribution of the corresponding relaxation rates,
weighted by the modes' contributions to the ISF.
A single mode means no memory; a broad spectrum means a decay that slows as the
fast modes die out. The moments of the spectrum give a hierarchy of successive
approximations, each a lower bound on the decay. In experiment the decay is
recorded only over a finite range of times, the time window, and a single
exponential is fitted over it. Fitted at late times, where only the slowest
modes remain, its rate, the dephasing rate, tends to the slowest relaxation rate, not to the
local rate; the
sum rule reaches the data instead through the area beneath the normalized decay, which
gives the correlation time. What the
general theory leaves open is everything that needs a model.

The present work supplies it for the Ising lattice gas, the simplest member of
the class that keeps the sign, strength and short-range correlations of the
interaction. Two adparticles on nearest-neighbour (NN) sites form a bond of
energy $-\varepsilon$, so that $\varepsilon>0$ is attraction and $\varepsilon<0$
repulsion. In the Kawasaki dynamics, an adparticle jumps to an empty NN site at a
rate lowered by the bonds the jump breaks.
The model has two control parameters, the surface coverage $\theta$ and the reduced
coupling $\beta\varepsilon$, with $\beta=1/k_BT$, $T$ the surface temperature and $k_B$
the Boltzmann constant. The value $\varepsilon=0$ defines the ideal layer, with
no interaction beyond the exclusion of double occupancy.
Two geometries are considered, because the experiment selects one. 
The momentum
transfer $\K$ lies along a crystal direction chosen in the measurement, the
azimuth, and the coherent
signal resolves only the jumps with a component along $\K$. On an anisotropic
surface the adparticles may be confined to channels, one-dimensional rows of
sites, and the measurement sees the projection of $\K$ on the channel.
On an isotropic surface the adlayer is two-dimensional, on the square or
triangular lattice of fcc(100) or fcc(111) faces. There the jumps orthogonal to
$\K$ drop out of the local and dephasing rates, but the correlations of the adlayer remain
two-dimensional. The channel is solved exactly and serves as a benchmark; the
surface requires approximations, which exact calculations on small clusters
test.

The theoretical treatment has two parts. In the statics, every coefficient of the decay is an
equilibrium average: the amplitude of the signal at $t=0$ is the static
structure factor (SSF) $F(\K)$, the Fourier transform of the pair correlation of
the occupations, and the rates are averages of the jump rates over the
equilibrium configurations. Along a channel, these averages are exact. On the
surface, they rest on the pair approximation, which treats every bond as if it
belonged to a chain, and on the random-phase approximation (RPA), which treats
every Fourier component of the occupations as an independent Gaussian mode. In
the dynamics, the effective hop rate $\Gamma_{\rm eff}$, the mean rate at which
an adparticle leaves its site, fixes with $F$ the local rate $\Omega(\K)$ and the
first order of the hierarchy, a single exponential. Along a channel, the dynamics
is, in addition, of gradient type,
where the net current between two NN sites can be seen as a current in a continuum following the difference of densities between two
points. The collective diffusion coefficient then follows exactly from
equilibrium averages. The gradient property is known in the mathematical theory
of lattice gases\cite{Spohn1991}; what it gives here is the memory of the
spin-echo signal in closed form. The channel thus serves as an exactly solvable
benchmark, and clusters of the square lattice show  the
asymmetry of the memory between attraction and repulsion. The memory is measured by the width of the relaxation
spectrum, $\Delta(\K)$, the variance of the rates relative to the square of their
mean, and by the coefficient $R=\Omega\tau_c$, with $\tau_c(\K)$ the correlation
time, the area beneath the decay divided by its initial value (normalized decay): $R=1$ for a single
exponential and $R>1$ when the memory slows the decay.

The paper establishes five results. The single exponential of the first order,
built from equilibrium quantities alone, reproduces the correlation time only at
weak coupling, nearly independently of coverage; only there can a measured
dephasing rate be read through the statics and the sum rule. The memory is
asymmetric: attraction builds it through the slow dissolution of clusters,
whereas a repulsive layer relaxes nearly as a single mode at its structure peak.
The dephasing rate, fitted over a time window where it is nearly constant,
approaches the slowest relaxation rate, not the local rate, so that only the area
beneath the decay gives the hop rate and the memory without fitting a rate. Along
a channel, the gradient property gives the collective diffusion coefficient
exactly and the width of the spectrum in closed form, and the static moments
alone tell whether the jump rates have this property. Finally, applied to
CO/Cu(001), the lattice gas reproduces the dephasing rates measured along two
azimuths; the data require a NN repulsion of at least $2$--$3\,k_BT$, whatever
the jump rates. The measured dephasing rate is then close to the inverse
correlation time, and the fitted model predicts how the dephasing rate, the
memory and the collective diffusion coefficient change with coverage.

Section~\ref{sec:theory} collects the results of the general theory that are
used. Section~\ref{sec:ising} introduces the lattice gas and its statics, and
Sec.~\ref{sec:dynamics} its dynamics. Section~\ref{sec:results} evaluates both
along a channel and on the surface. Section~\ref{sec:co} applies the model to the
measured rates of CO/Cu(001), tests the choice of jump rates and predicts the
coverage dependence.
Section~\ref{sec:concl} concludes and discusses the validity of the Ising
model, which requires well-defined adsorption sites and short-ranged pairwise
interactions and leaves the action of the interaction at the transition state of
a jump, the highest point of the path between two sites, to be chosen. Two
appendices give the finite-size convergence of the exact spectra and the third
moment of the spectrum along the channel.

\section{Theory}
\label{sec:theory}

The memory equation obeyed by the coherent intermediate scattering function
(ISF) of an interacting adlayer has been derived in ref.~\citen{PaperI},
together with the statements it allows about its coefficients. This section
states, without derivations, the results used in this work. None of them
depends on a particular form of the lateral interaction.

\subsection{The class and the signal}
\label{sec:cf}

The adlayer occupies the $N$ sites $\R$ of a Bravais lattice, with occupation
numbers $n_{\R}\in\{0,1\}$ and coverage $\theta$, the mean occupation per site.
The theory applies to the class of lattice gases defined in the Introduction:
the adparticles are conserved and jump to empty sites along a fixed set of jump
vectors $\mathbf L$, at jump rates $w_{\R\to\R+\mathbf L}$. These rates may depend on the
configuration and obey detailed balance, and the dynamics is invariant under the
lattice translations and inversion. Only the
convergence of the hierarchy of approximations
introduced below needs the rates to be of finite
range. Angular brackets $\avg{\cdots}$ denote averages over the equilibrium
distribution, unless stated otherwise.

The fluctuation of the occupation about its mean and its lattice Fourier
transform at the momentum transfer $\K$, the density wave, are
\begin{equation}
\delta n_{\R}\;\equiv\;n_{\R}-\theta ,
\label{eq:deltan}
\end{equation}
\begin{equation}
\delta n_{\K}\;=\;\sum_{\R}\delta n_{\R}\,e^{\,i\K\cdot\R} .
\label{eq:densitywave}
\end{equation}
respectively.
The mean square of the density wave per site,
\begin{equation}
S(\K)\;=\;\frac{\avg{|\delta n_{\K}|^{2}}}{N} ,
\label{eq:SKdef}
\end{equation}
measures the static fluctuations of the occupations at the wavelength
$2\pi/K$.

The ISF is built on the separation distribution
$\rho_{\R}(t)=\avg{n_{\mathbf 0}(0)\,n_{\R}(t)}/\theta$, the probability per
adparticle that a site at separation $\R$ is occupied at time $t$ given that the
origin was occupied at time $0$. Its lattice Fourier transform is the ISF. Since
the number of adparticles is conserved, $\rho_{\R}$ keeps its total weight, and
normalized by that weight its transform is a CF of the
separation.
The uniform part $\theta$ of $\rho_{\R}$ contributes only at the reciprocal
lattice vectors, where it gives the Bragg peaks of elastic scattering. At every
other momentum transfer, the ISF is the autocorrelation of one density wave,
$I(\K,t)=\avg{\delta n_{\K}^{*}(0)\,\delta n_{\K}(t)}/N\theta$. Its value at
$t=0$ is the SSF,
\begin{equation}
I(\K,0)\;=\;F(\K)\;=\;\sum_{\R}\bigl[P_{\R}-\theta\bigr]e^{\,i\K\cdot\R} ,
\label{eq:Fdef}
\end{equation}
with $P_{\R}=\rho_{\R}(0)$ the equilibrium pair distribution, the separation distribution at $t=0$, and it is
related to the fluctuation \eqref{eq:SKdef} by
\begin{equation}
F(\K)\;=\;\frac{S(\K)}{\theta} .
\label{eq:FSrel}
\end{equation}

The decay passes through three regimes: a ballistic one, much shorter than the
inverse friction coefficient of the substrate, in which the adparticle moves
freely within its well; the jump regime, in which the adparticles jump between
sites and the memory of earlier jumps acts; and the diffusive regime, much longer
than the mean residence time at a site, in which the decay approaches its slowest
rate. The theory describes the jump regime, continued back to $t=0$, and assumes
the ballistic regime and the equilibration inside a well to be short compared
with the residence time.

A spin-echo measurement records the decay over a finite range of times, the time
window, and is fitted there\cite{Jardine2009} to
\begin{equation}
I(\K,t)\;\simeq\;A(\K)+B(\K)\,e^{-\alpha(\K)\,t} ,
\label{eq:diffusive}
\end{equation}
with $A(\K)$ a static level, from immobile scatterers and any elastic
contribution, $B(\K)$ an amplitude and $\alpha(\K)$ the dephasing rate. The window
is usually chosen where the fitted rate is nearly constant. In energy, an
exponential decay $e^{-\alpha t}$ is a Lorentzian quasielastic line of full width
at half maximum $2\hbar\alpha$, with $\hbar$ the reduced Planck constant, and
spin-echo rates are often reported as this width. Neither $B$ nor $\alpha$ is
assumed here to be a property of the adlayer alone.

\subsection{The memory equation and the sum rule}
\label{sec:coeff}

The time evolution of any function of the configuration is generated by the
generator of the process, $\Lc$: acting on a function, it returns the sum over
all allowed jumps of the jump rate times the change the jump produces in that
function. Detailed balance makes $\Lc$ self-adjoint in the scalar product
defined by the equilibrium average\cite{liggett}, and $-\Lc$ non-negative.

Eliminating every variable of the adlayer but the density wave by the projection
of Nakajima and Zwanzig\cite{nakajima,zwanzig60} leaves at every momentum
transfer an exact equation of the form Mori gave to the motion of a single
variable\cite{Mori1965,Zwanzig2001},
\begin{equation}
\frac{\partial I(\K,t)}{\partial t}
\;=\;-\,\Omega(\K)\,I(\K,t)\;+\;\int_0^t\!dt'\;\Mc(\K,t-t')\,I(\K,t') ,
\qquad I(\K,0)=F(\K) .
\label{eq:eom}
\end{equation}
Here $\Omega(\K)$ is the local, or initial, rate, at which the decay of the jump regime sets
out, and $\Mc(\K,\tau)$ the memory function at delay $\tau$. Both are
equilibrium averages, and by the self-adjointness of $\Lc$ the memory function
is non-negative at every delay.

The local rate obeys the sum rule
\begin{equation}
\Omega(\K)\,F(\K)\;=\;\Gamma_{\rm eff}\,\bigl[\,1-f_{\K}\,\bigr] .
\label{eq:sumrule}
\end{equation}
Its scalar is fixed by the equilibrium jump frequencies, the mean number of jumps
per unit time along each jump vector,
\begin{equation}
J_{\mathbf L}\;=\;\bigl\langle\,w_{\R\to\R+\mathbf L}\,n_{\R}
\bigl(1-n_{\R+\mathbf L}\bigr)\bigr\rangle ,
\qquad
\Gamma_{\rm eff}\;=\;\frac{1}{\theta}\sum_{\mathbf L}J_{\mathbf L} ,
\label{eq:current}
\end{equation}
so that the effective hop rate $\Gamma_{\rm eff}$ is the mean rate at which one
adparticle leaves its site. Its momentum dependence is that of the single-jump
CF,
\begin{equation}
f_{\K}\;=\;\sum_{\mathbf L}p_{\mathbf L}\,e^{\,i\K\cdot\mathbf L} ,
\qquad
p_{\mathbf L}\;=\;\frac{J_{\mathbf L}}{\sum_{\mathbf L'}J_{\mathbf L'}} ,
\label{eq:fK}
\end{equation}
the CF of one jump vector drawn with the probabilities $p_{\mathbf L}$, the fraction of all jumps made along $\mathbf L$.

When the jump vectors are equivalent under the symmetry of the adlayer, as for
jumps to NN sites in a phase with the full symmetry of the
lattice, the weights are fixed by the lattice. The whole structure of the
adlayer then divides the rate, $\Omega=\Gamma_{\rm eff}[1-f_{\K}]/F$: the
narrowing of de Gennes\cite{deGennes1959} becomes a theorem on the lattice.
Jump vectors orthogonal to $\K$, for which $\cos(\K\cdot\mathbf L)=1$, drop out
of the product,
\begin{equation}
\Gamma_{\rm eff}\bigl[1-f_{\K}\bigr]\;=\;\frac1\theta\sum_{\K\cdot\mathbf L\neq0}
J_{\mathbf L}\bigl[1-\cos(\K\cdot\mathbf L)\bigr] ,
\label{eq:resolved}
\end{equation}
so that a measurement along $\K$ sees only the jump frequencies along the
directions it resolves; the orthogonal part of $\Gamma_{\rm eff}$ is cancelled by
that of $f_{\K}$. The orthogonal directions remain in the adlayer, however: the
SSF is the transform of the full pair correlation, to which separations
orthogonal to $\K$ contribute with unit phase, and the averages behind $\Omega$
and $\Mc$ are taken over the whole equilibrium layer.
At dilute coverage, the jumps are independent, the memory vanishes and
$F\to1$, and the decay is that of Chudley and Elliott\cite{Chudley1961},
\begin{equation}
I(\K,t)\;=\;e^{-\Gamma_0\,[1-f_{\K}]\,t} ,
\label{eq:ce}
\end{equation}
with $\Gamma_0$ the total hop rate of an isolated adparticle.
At long wavelength, the local rate becomes diffusive, $\Omega\to D_cK^{2}$, with
\begin{equation}
D_c\;=\;\frac{\Gamma_{\rm eff}\,\avg{\ell^{2}}}{2\,F(\mathbf 0)} ,
\qquad
\avg{\ell^{2}}\;=\;\sum_{\mathbf L}p_{\mathbf L}\bigl(\hat\K\cdot\mathbf L\bigr)^{2} ,
\label{eq:Dc1}
\end{equation}
$\avg{\ell^{2}}$ being the mean squared jump length projected on the direction
$\hat\K$ of the momentum transfer. This is a relation of Darken's
form\cite{Darken1948} for the coefficient of the local rate.

\subsection{The relaxation spectrum and the hierarchy}
\label{sec:modes}

Since $-\Lc$ is self-adjoint and non-negative, the decay is a superposition of
relaxation modes, each decaying exponentially at its relaxation rate
$\lambda_k(\K)\ge0$, the inverse of its relaxation time,
\begin{equation}
\frac{I(\K,t)}{F(\K)}\;=\;\sum_k a_k(\K)\,e^{-\lambda_k(\K)\,t} ,
\qquad a_k\ge0 ,\quad \sum_k a_k=1 .
\label{eq:modes}
\end{equation}
The rates weighted by $a_k$ form the relaxation spectrum $\mu_{\K}$, and the
average of a function $g$ of the rate over the spectrum is
$\avg{g(\lambda)}_{\mu}\equiv\sum_k a_k\,g(\lambda_k)$.
The moments of the spectrum, $m_n=\avg{\lambda^{n}}_{\mu}$, are static averages
of the adlayer, $m_n=\avg{\delta n_{\K}^{*}(-\Lc)^{n}\delta n_{\K}}/N\theta F$.
The first is the local rate, $m_1=\Omega$, whose value is given by the sum rule
\eqref{eq:sumrule}.
The second is
\begin{equation}
m_2(\K)\;=\;\frac{\avg{|\Lc\,\delta n_{\K}|^{2}}}{N\theta\,F(\K)} ,
\qquad
\Lc\,\delta n_{\K}\;=\;\sum_{\R,\mathbf L}
w_{\R\to\R+\mathbf L}\,n_{\R}\bigl(1-n_{\R+\mathbf L}\bigr)\,
e^{\,i\K\cdot\R}\bigl(e^{\,i\K\cdot\mathbf L}-1\bigr) ,
\label{eq:m2static}
\end{equation}
the lattice transform of the equilibrium correlation between jumps on two pairs
of sites.
A jump orthogonal to $\K$ leaves the density wave unchanged, so that $\Omega$ and
$m_2$ contain only the resolved jumps; the orthogonal motion enters the memory
from $m_3$ on, by changing the rates of the resolved jumps.

The memory starts at the spread of the rates,
\begin{equation}
\Mc(\K,0)\;=\;\Delta(\K)\,\Omega^{2}(\K) ,
\qquad
\Delta\;=\;\frac{m_2}{m_1^{2}}-1\;\ge\;0 ,
\label{eq:Delta}
\end{equation}
$\Delta$ being the variance of the spectrum relative to the square of its mean.
It vanishes if and only if $\delta n_{\K}$ is an eigenvector of $\Lc$. That is
the case when the jump rates are independent of the surrounding occupations, a
common total rate $\Gamma_0$ being shared equally among $z$ equivalent
neighbours: \eqref{eq:m2static} then reduces to
$\Lc\,\delta n_{\K}=-\Gamma_0[1-f_{\K}]\,\delta n_{\K}$. In this lattice gas
each adparticle jumps to an empty neighbouring site at a fixed rate, whatever
the other adparticles do, the only interaction being the exclusion of double
occupancy; it is known as the symmetric exclusion process\cite{Kutner1981},
symmetric because the jumps in opposite directions have equal rates. It relaxes through a single mode
at every coverage and momentum transfer, at the rate $\Gamma_0[1-f_{\K}]$ of the
isolated adparticle: its coherent rate does not depend on coverage, the site
blocking lowering $\Gamma_{\rm eff}$ and $F$ by the same factor $1-\theta$.
The moments generate Mori's continued fraction for the Laplace transform of the
decay\cite{Mori1965}. Its truncation at order $n$ uses the first $2n$ moments
and is the Gauss quadrature of $\mu_{\K}$, the approximation of the spectrum by
$n$ rates with positive weights that reproduces those moments. It gives a sum of
$n$ exponentials lying below the decay at every time and converging to it for an
interaction of finite range.
The first order is
\begin{equation}
I_1(\K,t)\;=\;F(\K)\,e^{-\Omega(\K)\,t} .
\label{eq:level1}
\end{equation}
Both its coefficients are exact, by \eqref{eq:Fdef} and \eqref{eq:sumrule}, and
it is a rigorous lower bound,
\begin{equation}
I(\K,t)\;\ge\;I_1(\K,t) ,
\label{eq:bound}
\end{equation}
which differs from the decay at short times by
\begin{equation}
\frac{I-I_1}{F}\;=\;\tfrac12\,\Delta\,(\Omega t)^{2}\,e^{-\Omega t}
+O\bigl((\Omega t)^{3}\bigr) ,
\label{eq:excess}
\end{equation}
so that \eqref{eq:level1} is exact wherever $\Delta$ vanishes. The second order
needs, besides $\Omega=m_1$, two coefficients fixed by the first three moments,
\begin{equation}
b_1\;=\;m_2-m_1^{2}\;=\;\Delta\,\Omega^{2} ,
\qquad
\Omega_1\;=\;\frac{m_3-2m_1m_2+m_1^{3}}{m_2-m_1^{2}} ,
\label{eq:coeffs1}
\end{equation}
and its memory function is the single exponential $b_1e^{-\Omega_1\tau}$.

\subsection{What a measurement returns}
\label{sec:measured}

The rate returned by the fit \eqref{eq:diffusive} depends on where the window
lies. At the origin, the decay sets out at the local rate $\Omega$, and at long
times it is carried by the slowest relaxation mode, of rate $\lambda_{\min}(\K)$
and weight $a_{\min}(\K)$ in \eqref{eq:modes}. A fit over a window returns an
effective rate between the two, close to $\Omega$ for an early window and
falling towards $\lambda_{\min}$ as the window moves to later times; it is nearly
constant only where one mode carries the signal. In the diffusive regime, the
dephasing rate and the amplitude tend to
\begin{equation}
\alpha(\K)\;=\;\lambda_{\min}(\K)\;\le\;\Omega(\K) ,
\qquad
B(\K)\;=\;a_{\min}(\K)\,F(\K)\;\le\;F(\K) ,
\label{eq:alphalmin}
\end{equation}
with equality only when one mode carries the signal, as it nearly does at long
wavelength. Where many relaxation rates lie close to $\lambda_{\min}$, the fitted
rate stays above $\lambda_{\min}$ and keeps falling slowly as the window moves.
The fitted pair is therefore not the pair of the first order, and
$B\alpha\le\Gamma_{\rm eff}[1-f_{\K}]$ is an inequality, not the sum rule.

The area beneath the normalized decay, the correlation time
$\tau_c=\int_0^{\infty}I\,dt/F=\avg{\lambda^{-1}}_{\mu}$, needs no time window. It
defines the coefficient
\begin{equation}
R(\K)\;\equiv\;\Omega(\K)\,\tau_c(\K)\;\ge\;1 ,
\qquad
\alpha\;\le\;\tau_c^{-1}\;\le\;\Omega ,
\label{eq:R}
\end{equation}
the factor by which the memory lengthens the correlation time beyond that of the
first order. The orders of the hierarchy approach it from below,
$1=R^{(1)}\le R^{(2)}\le\cdots\le R$, with
$R^{(2)}=[1-b_1/\Omega\Omega_1]^{-1}$.

In energy, the decay is seen as the quasielastic line or peak, the dynamic structure
factor $S(\K,\omega)=\pi^{-1}{\rm Re}\int_0^\infty I(\K,t)\,e^{i\omega t}dt$. By
\eqref{eq:modes} and \eqref{eq:eom},
\begin{equation}
S(\K,\omega)\;=\;\frac{F}{\pi}\sum_k\frac{a_k\,\lambda_k}{\lambda_k^{2}+\omega^{2}}
\;=\;\frac{F}{\pi}\,{\rm Re}\,\frac{1}{-i\omega+\Omega-\widetilde{\Mc}(-i\omega)} ,
\label{eq:Somega}
\end{equation}
with $\widetilde{\Mc}(s)=\int_0^\infty\Mc(\K,\tau)e^{-s\tau}d\tau$ the Laplace
transform of the memory function. The peak is a superposition of Lorentzians, one
for each relaxation mode, of half width $\hbar\lambda_k$ in energy; it is a single
Lorentzian, of full width at half maximum $2\hbar\Omega$, only without memory.
Its height at $\omega=0$ is $F\tau_c/\pi$, so that the peak value 
measures the correlation time.
The sum rule meets the data in the form
\begin{equation}
\frac{F(\K)}{\bigl[1-f_{\K}\bigr]\,\tau_c(\K)}\;=\;\frac{\Gamma_{\rm eff}}{R(\K)} .
\label{eq:Rmeas}
\end{equation}

A model must therefore supply the SSF, the jump frequencies \eqref{eq:current}
and the moments $m_2$ and $m_3$, all equilibrium averages. One property of the
jump rates matters in addition. The dynamics is of gradient
type\cite{Spohn1991} when the net current between two NN sites, averaged over
the jumps for a given configuration, behaves as a current in a continuum, which
follows the difference of densities between two points: it is the difference of
one local quantity, a function of the occupations around each site, taken at the
two sites. Then $R(\mathbf 0)=1$ and $D_c$ of \eqref{eq:Dc1} is exact at every
coupling.

\section{An Ising lattice-gas model}
\label{sec:ising}

\subsection{The lattice gas}
\label{sec:lg}

We consider the adlayer as a lattice gas, that is, as the Ising model written in
the variables natural to diffusion. Under the correspondence
$s_{\R}=2n_{\R}-1$\cite{Baxter1982} the Ising spin $s_{\R}=\pm1$ becomes the
occupation number $n_{\R}\in\{0,1\}$, the Ising exchange coupling becomes the
pairwise interaction between adparticles, and the magnetic field becomes the
chemical potential $\mu$, conjugate to the number of adparticles.

The interaction acts between nearest neighbours. The NN vectors
$\boldsymbol\delta$ of the lattice are its $z$ shortest non-zero lattice vectors,
$z$ being the coordination number, and the NN sites of a site $\R$ are the $z$
sites $\R+\boldsymbol\delta$. Two sites $\R$ and $\R+\boldsymbol\delta$ form an NN
pair; when both are occupied, their two adparticles form a bond, of energy
$-\varepsilon$, so that $\varepsilon>0$ is attraction and $\varepsilon<0$
repulsion. Summing over all NN pairs gives the grand-canonical Hamiltonian
\begin{equation}
	H \;=\; -\varepsilon \sum_{\langle \R\R'\rangle} n_{\R} n_{\R'} \;-\; \mu \sum_{\R} n_{\R} ,
	\label{eq:latticegas}
\end{equation}
where $\langle\R\R'\rangle$ runs over distinct NN pairs. The equilibrium
distribution of configurations is proportional to $e^{-\beta H}$, and the chemical potential is fixed by
$\avg{n_{\R}}=\theta\in(0,1)$. Everything below is therefore parameterized by the
pair $(\theta,\beta\varepsilon)$.

Throughout, the adlayer is taken in a uniform, disordered phase: every
equilibrium average is unchanged by the translations of the lattice and by its
rotations and reflections, so that $\avg{n_{\R}}=\theta$ at every site and the
$z$ NN vectors are equivalent. This excludes the ordered phases, in which the
occupation differs between sublattices, that strong repulsion produces, and the
coexistence of a dilute and a dense phase that strong attraction produces.

The statics must supply three kinds of equilibrium averages. The first is the
pair distribution $P_{\R}$, which builds the SSF \eqref{eq:Fdef}. The second
enters the hop rate. An adparticle at $\R$ can jump to an NN site
$\R+\boldsymbol\delta$, the target, only if the target is empty, and in the model
treated here its jump rate depends on the number
\begin{equation}
\nu_{\R}\;=\;\sum_{\boldsymbol\delta'\neq\boldsymbol\delta}n_{\R+\boldsymbol\delta'}
\label{eq:nu}
\end{equation}
of occupied NN sites of $\R$ other than the target, all $z$ of them if the target
is not an NN site, through the factor $e^{-\beta\varepsilon\nu_{\R}}$, the
Boltzmann factor of the $\nu_{\R}$ bonds that the jump breaks. The average needed
is therefore
\begin{equation}
\bigl\langle\,n_{\R}\bigl(1-n_{\R+\boldsymbol\delta}\bigr)\,e^{-\beta\varepsilon\,\nu_{\R}}\bigr\rangle
\;=\;\bigl\langle n_{\R}\bigl(1-n_{\R+\boldsymbol\delta}\bigr)\bigr\rangle\;
\bigl\langle e^{-\beta\varepsilon\nu_{\R}}\bigr\rangle_{\rm c} .
\label{eq:navg}
\end{equation}
The first factor is the probability that the departure site is occupied and the
target empty. The second is the generating function of $\nu_{\R}$ conditional on
that pair, the conditional average of any function $g$ of the configuration
being defined as
\begin{equation}
\avg{g}_{\rm c}\;\equiv\;
\frac{\bigl\langle n_{\R}\bigl(1-n_{\R+\boldsymbol\delta}\bigr)\,g\bigr\rangle}
{\bigl\langle n_{\R}\bigl(1-n_{\R+\boldsymbol\delta}\bigr)\bigr\rangle} .
\label{eq:condavg}
\end{equation}
The third kind consists of averages of products of the occupations of a few
neighbouring sites, which the moments $m_2$ and $m_3$ of Sec.~\ref{sec:modes}
require.

The two geometries of the Introduction differ in $z$. A one-dimensional channel
has $z=2$; it is realized on the stepped Cu(115) surface, where HeSE shows Na
atoms hopping along the step edges with collective correlations along the
steps\cite{Godsi2015}, and on the furrowed W(211) face, where field-ion
microscopy resolves a stronger interaction of Pd adatoms within a channel than
across channels\cite{Fu2004}. Along a channel the sites are labelled by an
integer $r$, with spacing $a$, so that the NN sites of $r$ are $r\pm1$. A momentum transfer $\K$ of modulus $K$, at an
angle $\tilde\beta$ to the channel, enters only through the phase
$x\equiv\Keff$ by which the density wave advances from one site to the next,
$e^{i\K\cdot\R_r}=e^{ixr}$; for $\K$ along the channel, $x=Ka$. The zone centre is
$x=0$ and the zone boundary (ZB) is $x=\pi$. The surface has $z=4$ on the square
lattice of fcc(100) faces, as for CO molecules on top of the Cu atoms of
Cu(001)\cite{Alexandrowicz2004}, and $z=6$ on the triangular lattice of fcc(111)
faces.

\subsection{The one-dimensional channel}
\label{sec:channel}

Along a channel the Boltzmann weight of a configuration factorizes into factors,
one for each NN pair $(r,r+1)$,
$\exp[\beta\varepsilon\sum_r n_rn_{r+1}+\beta\mu\sum_rn_r]=\prod_rT_{n_r,n_{r+1}}$,
with the transfer matrix\cite{Baxter1982}
\begin{equation}
	T_{n,n'} \;=\; e^{\,\beta\varepsilon\,n n'+\frac{\beta\mu}{2}(n+n')} ,
	\qquad n,n'\in\{0,1\} .
	\label{eq:Tdef}
\end{equation}
Its two eigenvalues $\Lambda_+>|\Lambda_-|$ are real and distinct. In the
thermodynamic limit, $N\to\infty$, the connected correlation of two sites, the
average product of their fluctuations $\delta n_r=n_r-\theta$, decays
geometrically with their separation $r$,
\begin{equation}
	\avg{\delta n_0\,\delta n_r} \;=\; \theta(1-\theta)\,\eta^{|r|},
	\qquad
	\eta \;\equiv\; \frac{\Lambda_-}{\Lambda_+}\in(-1,1),
	\label{eq:geometric}
\end{equation}
its amplitude being fixed at $r=0$ by $\avg{\delta n_0^{2}}=\theta(1-\theta)$;
for repulsion $\eta<0$ and the correlation alternates in sign from site to site.
The probabilities $p_{nn'}$ of the occupations $n,n'$ of two adjacent sites obey the
mass-action law
\begin{equation}
	\frac{p_{11}\,p_{00}}{p_{01}\,p_{10}} \;=\; \frac{T_{11}T_{00}}{T_{01}T_{10}} \;=\; e^{\beta\varepsilon} ,
	\label{eq:quasichem}
\end{equation}
exact for the chain, in which $\mu$ cancels. With the marginals
$p_{11}+p_{01}=\theta$, $p_{00}+p_{01}=1-\theta$ and, from \eqref{eq:geometric},
$p_{01}=\theta(1-\theta)(1-\eta)$, \eqref{eq:quasichem} is a quadratic whose
physical root is
\begin{equation}
	\eta(\theta,\beta\varepsilon) \;=\; \frac{\zeta-1}{\zeta+1},
	\qquad
	\zeta \;\equiv\; \sqrt{\,1-4\theta(1-\theta)\big(1-e^{\beta\varepsilon}\big)\,} .
	\label{eq:eta}
\end{equation}
This single number carries the whole correlation, with
$\eta\simeq\theta(1-\theta)\beta\varepsilon$ at weak coupling; since $|\eta|<1$
strictly, the chain has no phase transition.

The SSF follows from the same correlation. Conditioning \eqref{eq:geometric} on
an occupied origin gives the pair distribution $P_r=\theta+(1-\theta)\eta^{|r|}$.
Its uniform part $\theta$ gives the Bragg peaks, and its decaying part, by
\eqref{eq:Fdef}, a geometric series in the phase $x$ that sums to
\begin{equation}
	F(\mathbf{K}) \;=\; (1-\theta)\,\frac{1-\eta^{2}}{1-2\eta\cos x+\eta^{2}} .
	\label{eq:FK}
\end{equation}
This lattice Lorentzian has, at the zone centre and at the ZB, the values
\begin{equation}
	F(\mathbf{0})=(1-\theta)\,\frac{1+\eta}{1-\eta},
	\qquad
	F_{\rm ZB}=(1-\theta)\,\frac{1-\eta}{1+\eta},
	\label{eq:FKzone}
\end{equation}
reciprocal about the ideal value $1-\theta$: attraction piles intensity at the
zone centre and depletes the ZB, and repulsion does the reverse. The zone-centre
value is the compressibility of the adlayer, $F(\mathbf 0)=\partial\ln\theta/
\partial(\beta\mu)$, in units of its value for independent particles: a repulsive
layer is less compressible than the ideal one.

In the thermodynamic limit the occupations along the channel form a two-state
Markov chain in space: the joint probability of the occupations of $k$
consecutive sites is
\begin{equation}
P(n_1,\dots,n_k)\;=\;\pi_{n_1}\prod_{j=1}^{k-1}p(n_{j+1}|n_j) ,
\label{eq:markov}
\end{equation}
with $\pi_1=\theta$, $\pi_0=1-\theta$ and $p(n'|n)$ the probability that a site
has occupation $n'$ given that the preceding site has occupation $n$, the same
for the following site. These conditional probabilities are
\begin{equation}
\begin{gathered}
q\equiv p(1|1)=\theta+(1-\theta)\,\eta,
\qquad
p(0|1)=1-q=(1-\theta)(1-\eta),
\\
p(1|0)=\theta(1-\eta),
\qquad
p(0|0)=s\equiv1-\theta(1-\eta),
\end{gathered}
\label{eq:conditional}
\end{equation}
with $q$ the probability that a neighbour of an adparticle is occupied and $s$ the
probability that a neighbour of an empty site is empty. Every average over a few
consecutive sites is then a finite sum of products of \eqref{eq:conditional},
elementary in $\theta$ and $\eta$, and the occupations of the two NN sites of a
site are independent given the occupation of that site. The average
\eqref{eq:navg} is then exact; with $\nu_r$ \eqref{eq:nu} reduced to the
occupation of the NN site other than the target,
\begin{equation}
\bigl\langle n_r\bigl(1-n_{r\pm1}\bigr)\bigr\rangle=\theta(1-q),
\qquad
\avg{e^{-\beta\varepsilon\nu_r}}_{\rm c}=1-q+q\,e^{-\beta\varepsilon} .
\label{eq:navg1d}
\end{equation}

Finally, \eqref{eq:quasichem} with $p_{01}=\theta(1-q)$ writes the Boltzmann
factor of a bond in the same variables,
\begin{equation}
e^{-\beta\varepsilon}\;=\;\frac{\theta(1-\theta)(1-\eta)^{2}}{q\,s} ,
\label{eq:eboltz}
\end{equation}
so that every equilibrium average of the channel is a function of $\theta$ and
$\eta$ alone.

\subsection{The two-dimensional surface}
\label{sec:surface}

On the surface the partition function does not factorize. Two approximations are
used, one for the NN occupation statistics and one for the momentum dependence
of the SSF.

For the NN statistics we keep the $\eta(\theta,\beta\varepsilon)$ of
\eqref{eq:eta} as the correlation of every NN pair,
$\avg{n_{\mathbf 0}n_{\boldsymbol\delta}}=\theta^{2}+\theta(1-\theta)\eta$, with
$\boldsymbol\delta$ any of the $z$ NN vectors. This is the pair
approximation\cite{Pelizzola2005}: it treats each NN pair as if it belonged to
a chain and ignores the closed loops of the square or triangular lattice. It is
exact along a channel, accurate at weak coupling, at low coverage and near
saturation, and least accurate near half coverage.

The pair approximation also fixes the average \eqref{eq:navg}. Given that a
site is occupied, each of its NN sites is occupied independently with the
probability $q$ of \eqref{eq:conditional}. The target is then empty with
probability $1-q$, so that $\avg{n_{\R}(1-n_{\R+\boldsymbol\delta})}=\theta(1-q)$,
and the number $\nu_{\R}$ of occupied neighbours among the $z-1$ others is binomial,
with generating function
\begin{equation}
\avg{e^{-\beta\varepsilon\nu_{\R}}}_{\rm c}\;=\;\big[\,1-q+q\,e^{-\beta\varepsilon}\big]^{\,z-1} ,
\label{eq:mgf}
\end{equation}
which reduces to the exact \eqref{eq:navg1d} at $z=2$.

For the momentum dependence we use the RPA\cite{ChaikinLubensky,Hansen2013}, which treats each density wave
\eqref{eq:densitywave} as an independent Gaussian mode. The cost of a fluctuation
is the entropic term of an ideal layer plus the Fourier transform
$v(\K)=\sum_{\R}v(\R)e^{i\K\cdot\R}$ of the pair interaction energy $v(\R)$ of two
adparticles at separation $\R$, and equipartition gives, by \eqref{eq:FSrel},
\begin{equation}
	F(\mathbf{K}) \;=\; \frac{1-\theta}{1+\beta\,\theta(1-\theta)\,v(\mathbf{K})}.
	\label{eq:FKrpa}
\end{equation}
For the NN lattice gas, $v(\R)=-\varepsilon$ for each of the $z$ NN vectors and
zero otherwise, so that
\begin{equation}
	v(\mathbf{K}) \;=\; -\varepsilon\,z\,\gamma(\mathbf{K}),
	\qquad
	\gamma(\mathbf{K})=\frac{1}{z}\sum_{\boldsymbol\delta}e^{\,i\mathbf{K}\cdot\boldsymbol\delta},
	\label{eq:vtilde-nn}
\end{equation}
with $\gamma(\mathbf{K})$ the average of the phase factors $e^{i\K\cdot\boldsymbol\delta}$
over the NN vectors. It equals $1$ at the zone centre and has the minimum $-1$ on
a bipartite lattice, one that splits into
two sublattices whose sites have all their neighbours on the other, such as the
square lattice. On the square lattice
the minimum lies at the corner of the Brillouin zone, $\K=(\pi/a)(1,1)$ with $a$
the lattice spacing, where the density wave alternates between the two
sublattices, the $c(2\times2)$ modulation; a repulsive layer has its structure
peak there. The RPA SSF is then
\begin{equation}
	F(\mathbf{K}) \;=\; \frac{1-\theta}{1-\beta\varepsilon\,z\,\theta(1-\theta)\,\gamma(\mathbf{K})}.
	\label{eq:FK2d}
\end{equation}

At $z=2$, the RPA agrees with the exact \eqref{eq:FK} and \eqref{eq:FKzone}
only to leading order in $\beta\varepsilon$: at $\theta=0.5$ and
$\beta\varepsilon=-1.5$ the exact $F(\mathbf 0)$ is $0.236$ against $0.286$.
On the attractive side the RPA overestimates the growth of correlations, and
its denominator vanishes
at a coupling where the Gaussian treatment fails, an artefact of the
approximation, since the exact channel never diverges. Because it needs only $v(\mathbf{K})$, the RPA holds for an
interaction of any range, such as the oscillatory substrate-mediated
interaction of adsorbates on metals\cite{Repp2000}.

\section{Dynamics of the lattice gas}
\label{sec:dynamics}

This section supplies the dynamical inputs of Sec.~\ref{sec:theory} for the
lattice gas of Sec.~\ref{sec:ising}, at any coverage and coupling and on any
lattice. It defines the jump rates, derives the effective hop rate and the first
order of the hierarchy, and establishes the gradient property of the channel,
which makes the collective diffusion coefficient exact.

\subsection{The effective hop rate and the first order}
\label{sec:dyn}

The adparticles move by Kawasaki dynamics\cite{Kawasaki1966}: an adparticle
jumps to an empty NN site, which conserves the number of adparticles. The jump
crosses a transition state, the highest point of the path between the two sites,
which is taken as non-interacting: the adparticle there has no bonds. The
barrier of the jump therefore rises by $\nu_{\R}\varepsilon$ when $\nu_{\R}$ of
the NN sites of the departure site other than the target are occupied,
\eqref{eq:nu}, and the jump rate is
\begin{equation}
w_{\R\to\R+\boldsymbol\delta}\;=\;\frac{\Gamma_0}{z}\,e^{-\beta\varepsilon\,\nu_{\R}} ,
\label{eq:rate}
\end{equation}
with $\Gamma_0$ the total hop rate of an isolated adparticle, summed over its $z$
NN sites. The rate is set at the departure site, and the same form applies to
jumps to sites beyond the NN shell, with their own rate for an isolated
adparticle.

These rates obey detailed balance. For the reverse jump, from $\R+\boldsymbol\delta$
back to $\R$, the count \eqref{eq:nu} is $\nu_{\R+\boldsymbol\delta}$, the number
of occupied NN sites of the target other than the departure site. The jump breaks
$\nu_{\R}$ bonds and forms $\nu_{\R+\boldsymbol\delta}$, so that its energy change
is $\Delta H=-\varepsilon(\nu_{\R+\boldsymbol\delta}-\nu_{\R})$, while the reverse
jump breaks the $\nu_{\R+\boldsymbol\delta}$ bonds. Hence
\begin{equation}
\frac{w_{\R\to\R+\boldsymbol\delta}}{w_{\R+\boldsymbol\delta\to\R}}
\;=\;e^{-\beta\varepsilon(\nu_{\R}-\nu_{\R+\boldsymbol\delta})}\;=\;e^{-\beta\Delta H} ,
\label{eq:db}
\end{equation}
which is detailed balance with respect to the Hamiltonian \eqref{eq:latticegas}.
The rates and the statics $\eta$ and $q$ thus come from the same equilibrium
distribution.

The effective hop rate is the jump frequency \eqref{eq:current}, a sum of $z$
terms, equal in the uniform, disordered phase, each given by the average
\eqref{eq:navg},
\begin{equation*}
\Gamma_{\rm eff}\;=\;\frac{\Gamma_0}{\theta}\,
\bigl\langle\,n_{\R}\bigl(1-n_{\R+\boldsymbol\delta}\bigr)\,e^{-\beta\varepsilon\,\nu_{\R}}\bigr\rangle
\;=\;\Gamma_0\,(1-q)\,\bigl\langle e^{-\beta\varepsilon\nu_{\R}}\bigr\rangle_{\rm c} ,
\end{equation*}
the probability that the target of an adparticle is empty, times the
average Boltzmann factor of the bonds to be broken given that it is. The
interaction changes the bare rate in two ways. The target is empty with
probability $1-q=(1-\theta)(1-\eta)$ rather than $1-\theta$, which attraction
lowers and repulsion raises; and the bonds to the other occupied neighbours must
be broken, which \eqref{eq:mgf} averages. Hence
\begin{equation}
\Gamma_{\rm eff}(\theta,T)\;=\;\Gamma_0\,(1-\theta)(1-\eta)\,
\big[\,1-q+q\,e^{-\beta\varepsilon}\big]^{\,z-1},
\label{eq:gamma-eff}
\end{equation}
the total rate at which an adparticle leaves its site, valid for any lattice.
At $\varepsilon=0$ this is the site-blocking rate $\Gamma_0(1-\theta)$ of the
ideal layer, in which only the exclusion of double occupancy acts. Attraction
lowers it, both trapping the adparticle and blocking its target, and repulsion
raises it. Along a channel the two factors are exact, by the Markov property of
Sec.~\ref{sec:channel}, and with \eqref{eq:eboltz} the hop rate takes the compact
form
\begin{equation}
\Gamma_{\rm eff}\;=\;\Gamma_0\,\frac{(1-q)^{2}}{s}
\qquad(z=2) .
\label{eq:gamma-1d}
\end{equation}
On the surface \eqref{eq:gamma-eff} inherits the pair approximation of
\eqref{eq:mgf}. The bonds to all $z-1$ other neighbours change the hop rate, but
in the product $\Gamma_{\rm eff}[1-f_{\K}]$ the jump vectors orthogonal to $\K$
cancel, by \eqref{eq:resolved}. In the uniform, disordered phase the $z$ NN
vectors are equivalent, the probabilities $p_{\mathbf L}$ of \eqref{eq:fK} are
$1/z$, and $f_{\K}=\gamma(\K)$ of \eqref{eq:vtilde-nn}: the interaction
renormalizes the rate and not the geometry.

The density wave decays with amplitude $F(\K)$ and a local rate fixed by the sum
rule \eqref{eq:sumrule},
\begin{equation}
\Omega(\K)\;=\;\frac{\Gamma_{\rm eff}\,\big[1-f_{\K}\big]}{F(\K)} ,
\label{eq:alpha}
\end{equation}
the single-adparticle rate divided by the SSF: relaxation slows where the adlayer
has built up structure and quickens where structure is suppressed, the de Gennes
narrowing\cite{deGennes1959} in closed form. The first order of the hierarchy,
\eqref{eq:level1}, is then
\begin{equation}
I_1(\K,t)\;=\;F(\K)\,
\exp\!\left[-\,\frac{\Gamma_{\rm eff}\,\big[1-f_{\K}\big]}{F(\K)}\;t\right],
\label{eq:main}
\end{equation}
exact in amplitude and initial rate and a lower bound on the decay. At
$\varepsilon=0$ the site-blocking factor cancels between rate and amplitude and
\eqref{eq:main} is the exact decay of the symmetric exclusion process; as
$\theta\to0$ it returns the single-adsorbate theory\cite{TorresMiyares2026c};
and as $K\to0$ its rate tends to $D_cK^2$ with the coefficient \eqref{eq:Dc1}.

\subsection{Membership of the class and gradient dynamics}
\label{sec:member}

With the jump rates \eqref{eq:rate} the lattice gas \eqref{eq:latticegas} is a
member of the class of Sec.~\ref{sec:theory}. Its interaction and its jump rates
involve NN sites only, so that the hierarchy of Sec.~\ref{sec:modes} converges.
The moments $m_2$ and $m_3$, and whether the dynamics is of gradient type, are
fixed by the net currents between NN sites, since the density wave changes only
through them. Along a channel these currents are discrete gradients, as in a
continuum.

Along a channel the rate \eqref{eq:rate} of a jump from $r$ to $r\pm1$ depends
only on the occupation of $r\mp1$, the NN site of the departure site opposite
the target. Since an occupation takes the values $0$ and $1$, the rate is linear
in it,
\begin{equation}
w_{r\to r\pm1}\;=\;w_0+w_1\,n_{r\mp1},
\qquad
w_0=\frac{\Gamma_0}{2},\quad w_1=\frac{\Gamma_0}{2}\bigl(e^{-\beta\varepsilon}-1\bigr),
\label{eq:w01}
\end{equation}
$w_0$ being the rate of a jump to one side for an isolated adparticle and $w_1$
its change when the site behind is occupied. The mean current from $r$ to $r+1$
in a given configuration, the rate of jumps from $r$ to $r+1$ minus that of
jumps back, is the sum of $(w_0+w_1n_{r-1})n_r(1-n_{r+1})$ and
$-(w_0+w_1n_{r+2})n_{r+1}(1-n_r)$. Expanding the products, it becomes a
difference,
\begin{equation}
j_{r,r+1}\;=\;h_r-h_{r+1} ,
\qquad
h_r\;=\;n_r\bigl[\,w_0+w_1\bigl(n_{r-1}+n_{r+1}-n_{r-1}n_{r+1}\bigr)\bigr] ,
\label{eq:hfun}
\end{equation}
of one local quantity $h_r$, a function of the occupations of three consecutive
sites, the same at every site: the net current is a discrete gradient of $h$, and
the dynamics is of gradient type in the sense of
Sec.~\ref{sec:measured}\cite{Spohn1991,KipnisLandim1999}. With the fluctuation
$\delta h_r=h_r-\avg{h_r}$ and its lattice Fourier transform
$\delta h_{\K}=\sum_r\delta h_r\,e^{\,ixr}$, three consequences follow.

First, the generator acts on the density wave through $h$ alone. A jump
between $r$ and $r+1$ changes $n_r$ and $n_{r+1}$ by opposite units, so that
$\Lc\,n_r=j_{r-1,r}-j_{r,r+1}$, and with \eqref{eq:hfun}
\begin{equation}
\Lc\,\delta n_{\K}\;=\;-\,2\bigl(1-\cos x\bigr)\,\delta h_{\K} ,
\label{eq:Lgrad}
\end{equation}
exactly and at every momentum transfer.

Second, the moments are static correlations of $h$. Inserting
\eqref{eq:Lgrad} in \eqref{eq:m2static} gives
\begin{equation}
m_2(K)\;=\;\frac{4(1-\cos x)^{2}\,S_{hh}(K)}{\theta\,F(K)} ,
\qquad
S_{hh}(K)\;=\;\frac{\avg{|\delta h_{K}|^{2}}}{N} ,
\label{eq:Shh}
\end{equation}
with $S_{hh}$ the structure factor of $h$, as $S(K)$ of \eqref{eq:SKdef} is that
of the occupations. The correlation of $h$ at two sites is an average over a few
consecutive sites, which the Markov chain \eqref{eq:markov} evaluates; it
vanishes beyond NN sites, and with the $q$ and $s$ of \eqref{eq:conditional} the
relative width \eqref{eq:Delta} of the spectrum is
\begin{equation}
\Delta(K)\;=\;\frac{\eta^{2}}{1-2\eta\cos x+\eta^{2}}
\left[\frac{1+\eta}{1-q}+\frac{2s\,(1+\cos x)}{q}\right] .
\label{eq:Deltaclosed}
\end{equation}
It is of second order in $\eta$, whereas $F$ and $\Gamma_{\rm eff}$ change at
first order, and it is finite at long wavelength. In the same way $m_3$ is an
average over six consecutive sites, and the second order
of the hierarchy and $R^{(2)}$ are also in closed form along the channel.

Third, the collective diffusion coefficient is exact. A coverage that varies
slowly along the channel carries, in local equilibrium, the current
$-a\,\partial\avg{h_r}/\partial X$, $X$ being the position along the channel:
this is Fick's law with coefficient $a^{2}\,\partial\avg{h_r}/\partial\theta$.
Evaluated with the Markov chain, this coefficient equals
$\Gamma_{\rm eff}a^{2}/2F(0)$, and $R(\mathbf 0)=1$, so that
\begin{equation}
D_c\;=\;\frac{\Gamma_{\rm eff}\,a^{2}\cos^{2}\tilde\beta}{2\,F(0)}
\label{eq:Dcexact}
\end{equation}
holds along the channel exactly and at every coupling, as the coefficient of the
diffusive regime and not only of the local rate, the factor $\cos^{2}\tilde\beta$
projecting the channel on $\hat\K$.

The gradient property belongs to the jump rates, not to the lattice gas. The
Metropolis rate $w=\tfrac12\Gamma_0\min(1,e^{-\beta\Delta H})$ also obeys
detailed balance with respect to \eqref{eq:latticegas}, so that the statics are
unchanged and only the dynamics differs. Since it depends on the occupations
around both the departure site and the target, the mean current is no longer a
difference, its equilibrium fluctuation does not vanish at long wavelength, and
$m_2$ falls only as $K^{2}$, so that $\Delta$ diverges as $K^{-2}$. A finite
$\Delta(0)$ thus identifies gradient dynamics from static moments alone.

\section{Results}
\label{sec:results}

The closed forms of Secs.~\ref{sec:ising} and~\ref{sec:dynamics} are evaluated
along a channel ($z=2$) at $\theta=0.2$ for $\beta\varepsilon=-2,-1,0,+1,+2$, and
for the memory over the plane $0.1\le\theta\le0.9$,
$-2.5\le\beta\varepsilon\le2.5$. On the surface the square lattice ($z=4$) is
treated on clusters at $\theta=0.2$.

\subsection{Exact spectra}
\label{sec:spectrum}

The full relaxation spectrum of the channel is obtained exactly by applying the
generator of the jump rates \eqref{eq:rate} to a ring of $N$ sites in the
grand-canonical ensemble, in which the number of adparticles fluctuates and the
chemical potential is set to give the coverage of interest. The Lanczos
recursion builds, starting from the density wave, an orthonormal basis in the
equilibrium scalar product in which $\Lc$ is tridiagonal; its elements are the
coefficients $\Omega=\Omega_0,\Omega_1,\dots$ and $b_1,b_2,\dots$ of Mori's
continued fraction, the first of which appear in \eqref{eq:coeffs1}. The full
fraction gives the spectrum and the decay at every time. Its value at zero
frequency is the area beneath the normalized decay, the correlation time
$\tau_c=\int_0^\infty I\,dt/F$ of Sec.~\ref{sec:measured}, and with it $R$. Rings of
$N=14$ sites are used, and of $N=12$ for the map.

On the ring the sum rule \eqref{eq:sumrule} holds to machine precision with the
jump frequency \eqref{eq:current} of the ring itself. Its closed form
\eqref{eq:gamma-1d} agrees with it to within $10^{-5}$, the finite-size error of
the ring, which confirms that the hop rate is exact along the channel; so do
\eqref{eq:Lgrad}, \eqref{eq:Deltaclosed} and the closed-form
$R^{(2)}$. To compare with experiment, a single exponential
\eqref{eq:diffusive} is also fitted to the exact decay between $0.3F$ and
$0.05F$, which returns the pair $(B,\alpha)$ a measurement would report.

On the surface the same jump rates are applied to periodic clusters of
$L_\parallel\times L_\perp$ sites, $L_\parallel$ along $\K$ and $L_\perp$ across
it, up to twenty sites in all. The momentum transfers are those the cluster
admits: the point $X$, the ZB along a row of sites, $Ka=\pi$, and the zone corner
$M$, $\K a=(\pi,\pi)$, the wavevector of the $c(2\times2)$ modulation. With up
to $2^{20}$ configurations the continued fraction is not computed, and the same
correlation time is obtained directly. Since
$\int_0^\infty I\,dt=\avg{\delta n_{\K}^{*}(-\Lc)^{-1}\delta n_{\K}}/N\theta$,
it is
$\tau_c=\avg{\delta n_{\K}^{*}\psi}/\avg{|\delta n_{\K}|^{2}}$, where $\psi$ is
the solution of $-\Lc\,\psi=\delta n_{\K}$, found with the conjugate-gradient
method, an iterative solver for a self-adjoint non-negative operator.

\subsection{Structure and hop rate}
\label{sec:res-closed}

\begin{figure}[!t]
\centering
\includegraphics[width=0.99\textwidth]{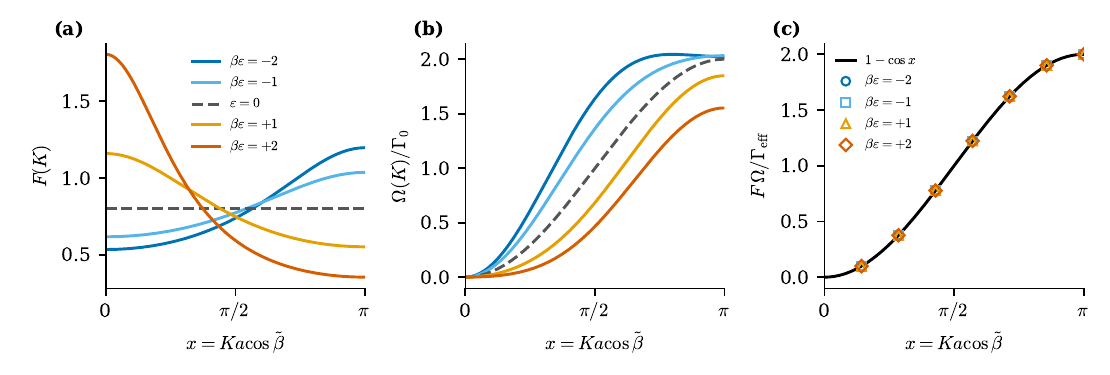}
\caption{Structure, local rate and sum rule along the channel ($z=2$).
(a) SSF $F(K)$, Eq.~\eqref{eq:FK}, at $\theta=0.2$ for
$\beta\varepsilon=-2$, $-1$ (blue), $0$ (dashed) and $+1$, $+2$ (orange).
(b) Local rate $\Omega(K)/\Gamma_0$, Eq.~\eqref{eq:alpha}.
(c) Sum rule: $F\Omega/\Gamma_{\rm eff}$ from the exact spectra of rings of
$N=14$ sites (symbols), divided by the closed-form hop rate
\eqref{eq:gamma-1d}, against $1-\cos x$ (line).}
\label{fig:statics}
\end{figure}

Figure~\ref{fig:statics} shows how the interaction shapes the two coefficients
of the first order \eqref{eq:main} along a channel, and that the sum rule ties
them together. Panels (a) and (b) give the SSF and the local rate across the
zone at $\theta=0.2$ for $\beta\varepsilon=-2$ to $+2$, and panel (c) tests the
sum rule against the exact spectra of the rings.

The SSF measures the static density fluctuations at the wavelength $2\pi/K$
[Fig.~\ref{fig:statics}(a)]. Attraction makes adparticles gather into clusters,
which enhances fluctuations at long wavelength and depletes the ZB, which
measures alternation from site to site. Repulsion favours that alternation and
keeps adparticles apart, so that the layer becomes less compressible and the
intensity moves to the ZB.

The local rate \eqref{eq:alpha} is the single-adparticle rate divided by the SSF,
and so mirrors it [Fig.~\ref{fig:statics}(b)]: a density wave relaxes slowly
where the layer holds it strongly in equilibrium. Attraction therefore slows the
long-wavelength relaxation, which requires the clusters to dissolve, and lowers
the whole curve through the smaller hop rate. Repulsion slows the relaxation near
the ZB, where the alternating arrangement persists; there the larger hop rate and
the larger $F_{\rm ZB}$ compensate, and $\Omega_{\rm ZB}$ stays close to its
ideal value $2\Gamma_0$.

The sum rule removes this structure [Fig.~\ref{fig:statics}(c)]. Along the
channel $f_K=\cos x$, and the product $F\Omega$ divided by the hop rate is
$1-\cos x$ at every coupling: the initial decay of a density wave is set by the
jumps out of each site, which see the interaction only through how often they
occur. The exact values of the rings, $F$ and $\Omega$ taken from the spectrum and
divided by the closed-form hop rate \eqref{eq:gamma-1d}, fall on this single
curve for the four non-zero couplings. The peaks and dips of panels (a) and (b) therefore
cancel exactly, and the coupling enters the product only through
$\Gamma_{\rm eff}$, which falls from $1.21\,\Gamma_0$ at $\beta\varepsilon=-2$ to
$0.28\,\Gamma_0$ at $+2$.

\subsection{The memory of the channel}
\label{sec:memory}

\begin{figure}[!t]
\centering
\includegraphics[width=0.75\textwidth]{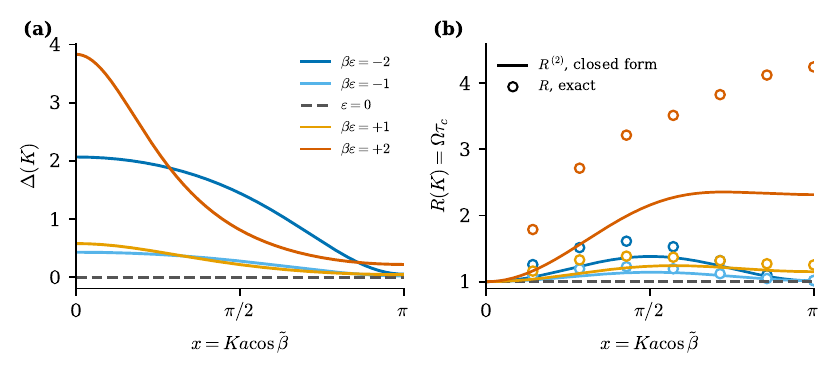}
\caption{The memory of the channel at $\theta=0.2$, for the couplings of
Fig.~\ref{fig:statics}.
(a) Width $\Delta(K)$, Eq.~\eqref{eq:Deltaclosed}.
(b) $R=\Omega\tau_c$, exact ($N=14$, circles), and $R^{(2)}$ in closed form
(lines).}
\label{fig:memory}
\end{figure}

The first order \eqref{eq:main} discards the memory, and the exact spectra of
the rings measure how much. Figure~\ref{fig:memory} gives the memory at $\theta=0.2$ for the couplings of
Fig.~\ref{fig:statics}. Panel (a) shows the relative width $\Delta$ of the
spectrum, which fixes how the decay sets out, \eqref{eq:excess}; and panel (b) the
coefficient $R$, by which the memory lengthens the correlation time, exact and in the
closed-form second order.

The width grows as $\eta^{2}$ and is concentrated at long wavelength
[Fig.~\ref{fig:memory}(a)]: at $\theta=0.2$, $\Delta(0)=0.43$ and $2.07$ for
$\beta\varepsilon=-1$ and $-2$, and $0.58$ and $3.83$ for $+1$ and $+2$, whereas at
the ZB it stays between $0.02$ and $0.22$. A long-wavelength density wave relaxes
only through many jumps, whose rates differ widely with the local arrangement,
bonded or free, while their mean, the local rate, is small; the spread relative
to the mean is then large. At the ZB a single jump already undoes the alternation,
and the rates are narrowly spread. By \eqref{eq:excess} the first order is
within ten per cent of the decay at $\Omega t=1$ wherever $\Delta\lesssim0.2$,
that is at the ZB for every coupling shown.

The coefficient $R$ tells a different story [Fig.~\ref{fig:memory}(b)], because it
weighs the slow rates, $R=\avg{\lambda}_\mu\avg{\lambda^{-1}}_\mu$, whereas
$\Delta$ weighs the fast ones. At $\beta\varepsilon=\pm1$ it stays below $1.23$
for repulsion and $1.39$ for attraction. At $-2$ it reaches $1.61$ at mid-zone and
returns to $1.02$ at the ZB, where the alternating arrangement of the repulsive
layer relaxes nearly as a single mode. At $+2$ it grows to $4.24$ at the ZB,
although $\Delta$ is there only $0.22$: the dissolution of clusters is a slow
process, which carries a fifth of the signal at a small fraction of the local
rate. The memory thus acts on the slow tail of the decay, and far more strongly
for attraction than for repulsion. The closed-form second order captures most of
it at moderate coupling, $R^{(2)}\le1.14$ and $1.24$ at $\beta\varepsilon=-1$ and
$+1$ against $1.23$ and $1.39$, but only about half at $+2$, where the higher
orders recover the rest from below, $R^{(5)}=3.69$ against $R=4.24$ at the ZB.

The rings also confirm the gradient signature of Sec.~\ref{sec:member} at small
momentum transfer. On rings of $N=8$ to $18$ sites, at $\theta=0.2$ and
$\beta\varepsilon=+1.5$, $\Delta(K)(Ka)^{2}$ falls from $0.67$ to $0.18$ between
$Ka=\pi/4$ and $\pi/9$ for the jump rates \eqref{eq:rate}, as a finite $\Delta(0)$
requires, but only from $0.52$ to $0.27$ for a Metropolis rate, levelling off as a
divergence $\Delta\propto K^{-2}$ requires; at $\theta=0.5$, $\beta\varepsilon=-1.5$
the two fall from $0.21$ to $0.04$ and from $0.08$ to $0.05$.

\begin{figure}[!t]
\centering
\includegraphics[width=0.56\textwidth]{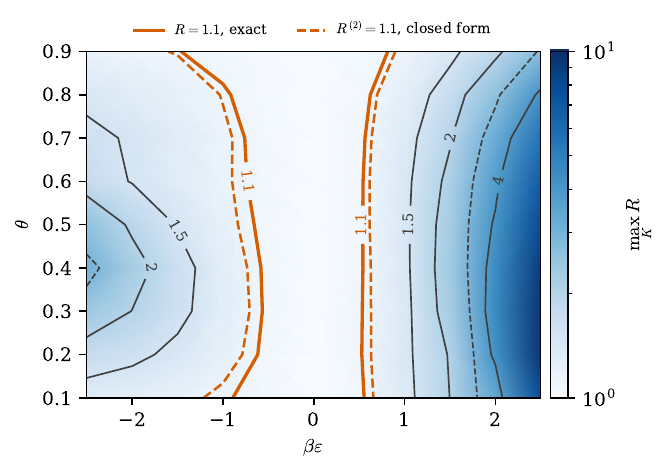}
\caption{Largest value across the zone of the exact $R=\Omega\tau_c$ along the
channel ($N=12$, at most six per cent low at the strongly attractive corner),
with its contours (solid) and those of the closed-form $R^{(2)}$ at $1.1$ and $2$
(dashed). The contour $R=1.1$ bounds the region where the first order
\eqref{eq:main} gives the correlation time to within ten per cent.}
\label{fig:map}
\end{figure}

Figure~\ref{fig:map} shows where in the plane of coverage and coupling the first
order can be used. It gives the largest exact $R$ across the zone, with the
contour $R=1.1$, which marks a correlation time within ten per cent of the first
order, and the same contours of the closed-form $R^{(2)}$. The region where the
first order holds is only a band around the ideal layer,
$-0.8\lesssim\beta\varepsilon\lesssim0.55$, nearly independent of coverage. Beyond
it the memory grows asymmetrically. On the attractive side clusters form at every
coverage, and $R$ exceeds $2$ at $\beta\varepsilon=1.5$ up to $\theta\approx0.6$
and reaches $9$ at $\beta\varepsilon=2.5$. On the repulsive side it exceeds $2$
only for $\beta\varepsilon\lesssim-2$ near $\theta\approx0.4$, where the
alternating arrangement is most developed. The closed-form $R^{(2)}$ reproduces
the shape of the map with every contour shifted outward, as a lower bound
requires, so that it locates the region of validity of the first order without
the exact spectrum.

\begin{figure}[!t]
\centering
\includegraphics[width=0.7\textwidth]{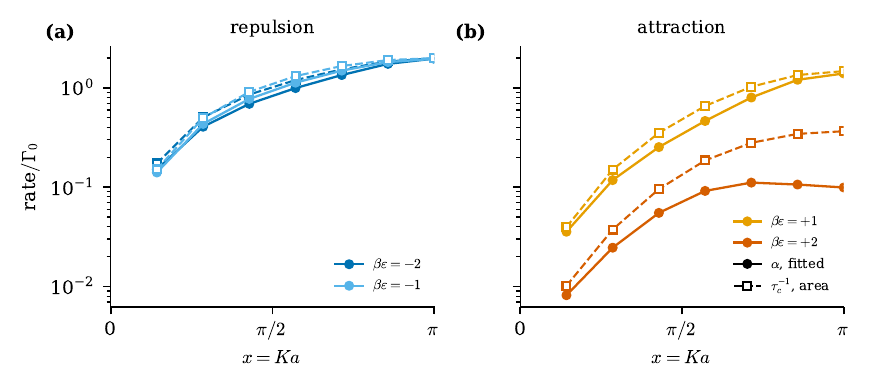}
\caption{Fitted rate $\alpha$ (filled symbols, solid) and inverse correlation
time $\tau_c^{-1}$ (open symbols, dashed), in units of $\Gamma_0$, across the zone
at $\theta=0.2$, for (a) repulsion and (b) attraction, from the exact decays of
rings of $N=14$ sites. The fit is a single exponential between $0.3F$ and
$0.05F$, which corresponds to times from $(1.0$--$1.2)\,\tau_c$ to
$(2.9$--$3.5)\,\tau_c$, except at $\beta\varepsilon=+2$ beyond mid-zone, where it
extends from $0.4\,\tau_c$ to $5.3\,\tau_c$; in units of $\Gamma_0^{-1}$ the
window runs from $0.6$--$1.5$ at the repulsive ZB to $110$--$330$ at the smallest
momentum transfer for $\beta\varepsilon=+2$. The correlation time $\tau_c$ is the area beneath the entire decay and needs no fit. The fitted
$\alpha$ stays below $\tau_c^{-1}$, as $\lambda_{\min}$ does by \eqref{eq:R},
with equality only for a single mode.}
\label{fig:window}
\end{figure}

Figure~\ref{fig:window} shows what this memory does to a measurement. It
compares, across the zone at $\theta=0.2$, the rate $\alpha$ of a single
exponential fitted to the exact decay between $0.3F$ and $0.05F$, a late window
of the kind used in measurements, with the inverse correlation time
$\tau_c^{-1}$, which needs no fit; panel (a) shows repulsion and panel (b)
attraction.

At long wavelength this window lies many residence times $1/\Gamma_{\rm eff}$
out, where one mode carries nearly the whole signal, and $\alpha=\lambda_{\min}$.
Towards the ZB it spans about one residence time, and there the slowest mode
carries almost no weight, $10^{-2}$ or less on rings of $12$ to $18$ sites and
decreasing with the size: the spectrum becomes continuous at its slow end, and
$\alpha$ is an effective rate of the window. The fitted rates stay below
$\tau_c^{-1}$ at every momentum transfer, as \eqref{eq:R} requires of
$\lambda_{\min}$. For repulsion they nearly coincide
[Fig.~\ref{fig:window}(a)]: $\alpha\tau_c=0.81$--$0.99$, closest to unity at the
ZB, where the alternating arrangement relaxes nearly as a single mode. For
attraction they separate [Fig.~\ref{fig:window}(b)]. At $\beta\varepsilon=+1$
$\alpha\tau_c$ falls to $0.71$ at mid-zone, and at $+2$ the fitted rate stops
growing beyond mid-zone, at about $0.1\,\Gamma_0$, while $\tau_c^{-1}$ keeps
rising to $0.37\,\Gamma_0$ at the ZB, so that $\alpha\tau_c$ falls to $0.27$: the
fit sees only the slow dissolution of clusters, whereas the area also counts the
fast relaxation that precedes it. The fitted product inherits the deficit and lies
below the window-free $\Gamma_{\rm eff}/R$ of \eqref{eq:Rmeas}, most for
attraction.

In practice the fit is moved to later times until the rate stops
varying\cite{TownsendThesis}; applied to the exact decay, this lowers $\alpha$
steadily, with no plateau before the signal is exhausted, so that the plateau
sought in practice is set by the noise, as a drift seen in measured decays
suggests\cite{TownsendThesis}. The area, and with it $\tau_c$, depends on no such
choice.

\subsection{The memory on the surface}
\label{sec:memory2d}

On the surface the closed forms carry the pair approximation, and the memory can
be computed only on small clusters. Table~\ref{tab:2d} collects what the clusters
of the square lattice give at $\theta=0.2$, at the point $X$. For each coupling it
lists the SSF and the relative width $\Delta$ at that point, and the exact $R$ on
clusters of increasing transverse size, with its extrapolation to infinite size.

On every cluster the sum rule \eqref{eq:sumrule}, with the jump frequencies of the
resolved directions only, \eqref{eq:resolved}, holds to machine precision: the
jumps orthogonal to $\K$ cancel between $\Gamma_{\rm eff}$ and $f_{\K}$ although
they are correlated with the resolved ones. The pair-approximation hop rate
\eqref{eq:gamma-eff} is within two per cent of the exact one at
$|\beta\varepsilon|\le1$.

\begin{table}[!h]
\centering
\caption{Exact $R=\Omega\tau_c$ on the square lattice at $\theta=0.2$, at the
point $X$, on clusters of transverse size $L_\perp=3$, $4$, $5$ (longitudinal
size $4$), with its geometric extrapolation $R_\infty$; $F$ and $\Delta$ from the
$4\times4$ cluster. The value marked $\sim$ is not converged.}
\label{tab:2d}
\vspace{4pt}
\small
\begin{tabular}{c cc ccc c}
\hline
$\beta\varepsilon$ & $F$ & $\Delta$ & $L_\perp=3$ & $4$ & $5$ & $R_\infty$\\
\hline
$-2$   & $0.68$ & $4.55$ & $2.23$ & $1.73$ & $1.76$ & $1.75$\\
$-1$   & $0.75$ & $0.56$ & $1.34$ & $1.27$ & $1.26$ & $1.26$\\
$+1$   & $0.68$ & $0.55$ & $1.92$ & $1.65$ & $1.56$ & $1.50$\\
$+1.5$ & $0.42$ & $1.20$ & $6.43$ & $5.13$ & $4.40$ & $\sim3.4$\\
\hline
\end{tabular}
\end{table}

At weak and moderate coupling the memory on the surface is comparable to that
along the channel, though somewhat larger (Table~\ref{tab:2d}): $R\approx1.26$ at
$\beta\varepsilon=-1$ and $1.50$ at $+1$, against at most $1.23$ and $1.39$ along
the channel. The surface adds two sources of memory. An adparticle has more
neighbours, so that its jump rate takes more values and the spectrum is wider;
and the jumps orthogonal to $\K$, absent from $\Omega$ and $m_2$, enter the memory
from $m_3$ on (Sec.~\ref{sec:modes}).

For repulsion the memory depends on whether the momentum transfer matches the
order towards which the layer tends. On the square lattice repulsion favours the
$c(2\times2)$ arrangement, whose wavevector is the zone corner $M$, not the point
$X$ of the table. On the $4\times4$ cluster $R=1.00$ at $M$ for
$\beta\varepsilon=-1$, and at $\beta\varepsilon=-2$, $R\approx1.75$ at $X$ but $1.21$ at $M$, just
as the channel relaxes nearly as a single mode, $R=1.02$, at its own ordering
wavevector, the ZB. At the structure peak of a repulsive layer the arrangement
it favours is restored by single jumps, and the memory lies away from it.

For attraction the surface differs qualitatively from the channel: it condenses.
At $\theta=0.2$ the coexistence of a dilute and a dense phase is reached at
$\beta\varepsilon=1.77$\cite{Baxter1982}. As this coupling is approached the
clusters grow and their dissolution slows, and the memory grows steeply, to
$R\approx3$ or more at $\beta\varepsilon=1.5$. There the clusters exceed the size
of the computed ones, $R$ still changes strongly with the transverse size, and
the value marked $\sim$ is indicative only.

\section{Application to CO/Cu(001)}
\label{sec:co}

CO on Cu(001) is a system to which the NN lattice gas applies. The molecules sit
on top of the Cu atoms, a square lattice of spacing $a=2.55$~\AA, and saturate
at $\theta=0.5$ in a $c(2\times2)$ structure, in which no two molecules occupy
NN sites\cite{Alexandrowicz2004}: the NN interaction is repulsive. The barrier
for diffusion is about $125$~meV\cite{Alexandrowicz2004,Jardine2004}, nearly
eight times $k_BT=16.4$~meV at $T=190$~K, so that the molecules move by
well-separated jumps, the jump regime of Sec.~\ref{sec:cf}.

The experiment\cite{Alexandrowicz2004} fits a single exponential
\eqref{eq:diffusive} to the polarization over spin-echo times up to $0.45$~ns, at
$\theta=0.1$, along the $\langle110\rangle$ azimuth, which runs along the NN
vectors, and the $\langle100\rangle$ azimuth, along the diagonals. It reports the
dephasing rate $\alpha$ as the width $2\hbar\alpha$ of the quasielastic line
\eqref{eq:Somega} that a single exponential would give. We convert the $56$
published widths, with their error bars, back to $\alpha(\K)$, with
$2\hbar=1.316~\mu$eV\,ns. This $\alpha$ is meant to be the slowest
relaxation rate; by Sec.~\ref{sec:measured} a fit over a finite window returns
it only where one mode carries the decay, and otherwise an effective rate above
it.
On $\langle110\rangle$ the zone boundary is the point $X$, at
$\Delta K=\pi/a=1.23$~\AA$^{-1}$; on $\langle100\rangle$ it is the zone corner
$M$ of Sec.~\ref{sec:surface}, at $\Delta K=\sqrt2\,\pi/a=1.74$~\AA$^{-1}$, the
wavevector of the $c(2\times2)$ arrangement.

A molecule jumps in two ways: to an empty NN site across the bridge site between
two Cu atoms, a jump labelled B (bridge), and to an empty diagonal site across
the fourfold hollow site, a jump labelled H (hollow). Both follow the rates
\eqref{eq:rate}: a molecule with $\nu_{\R}$ occupied NN sites \eqref{eq:nu}
makes each of its four B jumps at rate $w_Be^{-\beta\varepsilon\nu_{\R}}$ and each
of its four H jumps at rate $w_He^{-\beta\varepsilon\nu_{\R}}$, with $w_B$ and
$w_H$ the rates of an isolated molecule, which jumps at the total rate
$\Gamma_0=4(w_B+w_H)$. The model has three parameters, $\Gamma_0$, $w_H/w_B$ and
$\beta\varepsilon$. For given parameters, kinetic Monte Carlo on a $40\times40$
lattice with periodic boundaries samples the exact equilibrium of the layer and
gives the ISF along both azimuths. From it we obtain $\alpha$ exactly as the
experiment does, by fitting a single exponential over the first $0.45$~ns, and
also $F$, the jump frequencies \eqref{eq:current} and hence $\Omega$ from the sum
rule, $\tau_c$ from the area, and $\Delta$ from $m_2$ \eqref{eq:m2static}
evaluated on the sampled configurations.

The agreement between model and data is measured by
$\chi^2=\sum_i(\alpha_i-\alpha_i^{\rm model})^2/\sigma_i^2$, the sum over the
$56$ measured rates $\alpha_i$, with error bars $\sigma_i$, of the squared
deviations from the model in units of the error bars. For a model that describes
the data, and error bars that describe their scatter, $\chi^2$ is about the
number of points less the number of fitted parameters. For each coupling,
$\Gamma_0$ and $w_H/w_B$ are chosen to minimize $\chi^2$, which then measures how
well that coupling can describe the data.

\begin{figure}[!t]
\centering
\includegraphics[width=0.99\textwidth]{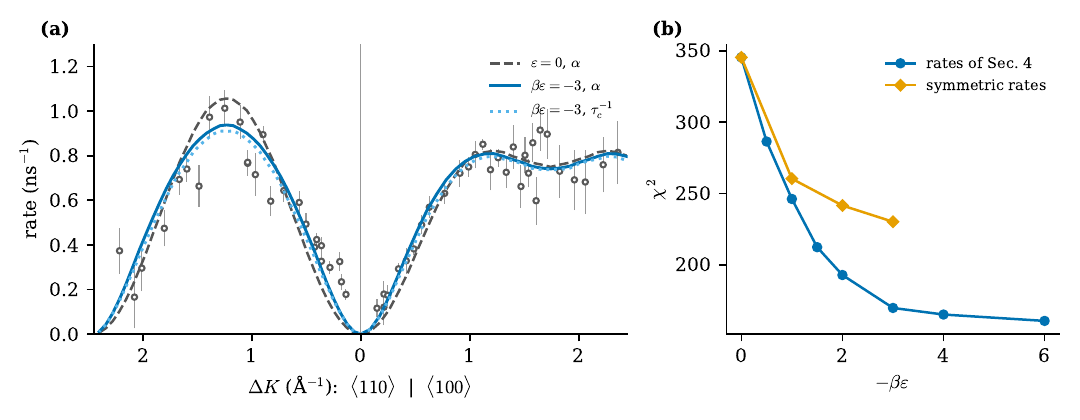}
\caption{The lattice gas fitted to CO/Cu(001) at $\theta=0.1$, $T=190$~K.
(a) Measured dephasing rate $\alpha$ along $\langle110\rangle$ (left) and
$\langle100\rangle$ (right)\cite{Alexandrowicz2004} (circles), with the model
$\alpha$ fitted over $0.45$~ns for $\varepsilon=0$ (dashed grey) and
$\beta\varepsilon=-3$ (solid), and for the latter $\tau_c^{-1}$ (dotted).
(b) $\chi^2$ against the coupling, $\Gamma_0$ and $w_H/w_B$ refitted at each
point, for the rates \eqref{eq:rate} (circles) and for symmetric rates
(diamonds).}
\label{fig:cofit}
\end{figure}

Figure~\ref{fig:cofit} shows the fit and what it implies at the measured
coverage. Panel (a) compares the measured $\alpha$ with the model $\alpha$ for
the ideal layer and for a strong repulsion, $\beta\varepsilon=-3$, and gives for
the latter the inverse correlation time $\tau_c^{-1}$. Panel (b) gives the
minimal $\chi^2$ against the coupling, for the rates \eqref{eq:rate} and for
symmetric rates, defined below.

The data select a repulsive layer [Fig.~\ref{fig:cofit}(b)]: $\chi^2$ falls from
$345$ for the ideal layer to $193$ at $\beta\varepsilon=-2$ and levels off beyond
$-3$, at $161$ when NN pairs are excluded altogether. They thus require a
repulsion of at least $2$--$3\,k_BT$, about $35$--$50$~meV, and bound it only from
below: at $190$~K NN pairs are nearly absent, as the $c(2\times2)$ saturation
implies. At $\beta\varepsilon=-3$ the fit gives $\Gamma_0=0.60$~ns$^{-1}$ and
$w_H/w_B=0.52$, which with equal attempt frequencies places the hollow barrier
about $11$~meV above the bridge barrier. The sum rule explains the gain. It comes
at long wavelength, where on both azimuths the measured $\alpha$ exceeds that of
the ideal layer [Fig.~\ref{fig:cofit}(a)]: a repulsive layer is less
compressible, $F=0.60$ at the smallest momentum transfer against $0.90$ for the
ideal layer, and by \eqref{eq:alpha} its rate is larger there by the same factor.
The remaining $\chi^2$, three per point, reflects a scatter of the data beyond
their error bars; half of it comes from the ten points on $\langle110\rangle$
below $0.6$~\AA$^{-1}$, which lie above the model. Changing the absolute scale of
the published widths, which the original analysis notes is uncertain by a factor
of order two\cite{Alexandrowicz2004}, by factors of $0.6$ or $1.85$ changes $\chi^2$ by at
most $15$ and leaves its dependence on the coupling unchanged.

The rates \eqref{eq:rate} assume a transition state without bonds. With
symmetric rates instead, $w_{B,H}\,e^{-\beta\Delta H/2}$, in which the barrier
changes by half the energy change $\Delta H$ of the jump, the repulsion is still
required, $\chi^2$ falling to $230$ at $\beta\varepsilon=-3$: it is a property of
the data, not of the choice of rates.

The dephasing rate is close to the inverse correlation time: the model $\alpha$
and $\tau_c^{-1}$ nearly coincide across the zone [Fig.~\ref{fig:cofit}(a)], so
that the measured rate reflects the area beneath the decay. The reason lies in
the decay itself. With the fitted parameters, beyond its first
instants the ISF is nearly a single exponential at every momentum transfer, so
that the exponential fitted over the experimental window follows the slow decay.
At $X$ the decay is initially faster: within the first tenth of a nanosecond the few
NN pairs break, since a pair forms at the rate of an isolated molecule but, by
\eqref{eq:rate}, breaks faster by the factor $e^{-\beta\varepsilon}$. This fast
part carries little weight, and the fitted $\alpha=0.94$~ns$^{-1}$ at $X$ is within
$4\%$ of $\tau_c^{-1}=0.91$~ns$^{-1}$. At $M$ the decay is a single exponential to
within $3\%$, $R=1.03$: the $c(2\times2)$ modulation is the structure peak of the
layer and is restored by single jumps, as on the clusters of
Sec.~\ref{sec:memory2d}.

\begin{figure}[!t]
\centering
\includegraphics[width=0.99\textwidth]{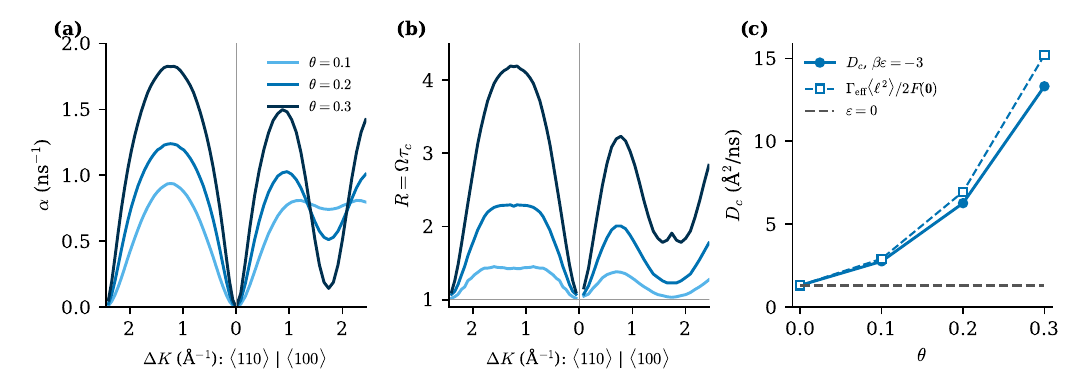}
\caption{Dephasing rate and memory of CO/Cu(001) from the fitted model
($\beta\varepsilon=-3$, $w_H/w_B=0.52$, $\Gamma_0=0.60$~ns$^{-1}$) along
$\langle110\rangle$ (left) and $\langle100\rangle$ (right), at $\theta=0.1$, $0.2$
and $0.3$. (a) $\alpha$ fitted over $0.45$~ns, (b) $R=\Omega\tau_c$. (c) Collective
diffusion coefficient $D_c$ against
coverage (circles), Darken form $\Gamma_{\rm eff}\avg{\ell^{2}}/2F(\mathbf 0)$ of
\eqref{eq:Dc1} (squares) and ideal layer (dashed).}
\label{fig:comem}
\end{figure}

Figure~\ref{fig:comem} follows the dephasing rate, the memory and the collective
diffusion coefficient to higher coverage, as a prediction of the fitted model.
Panels (a) and (b) give $\alpha$ fitted over $0.45$~ns and $R=\Omega\tau_c$
across the zone at $\theta=0.1$, $0.2$ and $0.3$; panel (c) gives $D_c$
against coverage, obtained from $\alpha/K^{2}$ at the smallest momentum transfer
of the simulation, $0.06$~\AA$^{-1}$, together with the Darken form
\eqref{eq:Dc1} and the value of the ideal layer.

At the measured coverage the memory is moderate, $R\le1.43$
[Fig.~\ref{fig:comem}(b)], and the two measures of it behave differently.
$\Delta$, which weighs the fast rates, is large, $29$ at $X$ and $8$ at $M$ for
$\beta\varepsilon=-3$: it is set by the rare NN pairs and their fast
break-up, and it grows rapidly with the repulsion, from $3.4$ at $X$ for
$\beta\varepsilon=-2$ to $71$ for $-4$, so that the data, which bound the
repulsion only from below, do not fix it. On the square lattice these rates are
not of gradient type, and $\Delta$ also rises at long wavelength, as
Sec.~\ref{sec:member} found for non-gradient rates. $R$ and $\tau_c$, which weigh
the slow rates, are robust: between $\beta\varepsilon=-2$ and $-6$, $R$ at $X$
stays between $1.33$ and $1.47$ and $\tau_c$ changes by less than $3\%$. The
correlation time, and with it what a fit returns, is thus fixed by the data,
while the initial spread of the decay is not.

In the ideal layer the coherent rate does not depend on coverage, the site
blocking cancelling between $\Gamma_{\rm eff}$ and $F$ (Sec.~\ref{sec:modes}); it
thus needs no site-blocking correction, unlike the reduction by a factor of $1.5$
applied to the simulated rates in the original analysis\cite{Alexandrowicz2004}.
The repulsive layer behaves otherwise. On $\langle110\rangle$ the dephasing rate
at $X$ rises from $0.94$ to $1.24$ and $1.82$~ns$^{-1}$ at $\theta=0.2$ and $0.3$
[Fig.~\ref{fig:comem}(a)], while $R$ grows to $2.3$ and $4.2$
[Fig.~\ref{fig:comem}(b)]: the more NN pairs the layer holds, the larger the fast
part of the decay, which the dephasing rate does not see. On $\langle100\rangle$ the rate
at $M$ falls instead, to $0.51$ and $0.14$~ns$^{-1}$, because $F(M)$ grows, to $2.7$
and $10.6$, as the layer builds up $c(2\times2)$ short-range order: this is the
narrowing \eqref{eq:alpha} at the structure peak, with $R$ rising to $1.9$ at
$\theta=0.3$ as that order becomes slow to rearrange. The maximum of $\alpha$ on
$\langle100\rangle$ moves away from $M$, and the two azimuths, nearly equal at
$\theta=0.1$, separate. At $\theta=0.3$ and $\beta\varepsilon=-3$ the layer is
close to $c(2\times2)$ order, and these values are the most sensitive to the
strength of the repulsion, which the data bound only from below.

The collective diffusion coefficient shows the same physics at long wavelength
[Fig.~\ref{fig:comem}(c)]. In the ideal layer it does not depend on coverage,
$D_c=1.31$~\AA$^2$/ns with the fitted $\Gamma_0$. The repulsive layer diffuses
faster as it fills: $D_c=2.8$, $6.3$ and $13.3$~\AA$^2$/ns at $\theta=0.1$, $0.2$
and $0.3$, because the hop rate rises, molecules with occupied neighbours leaving
their sites faster, and $F(\mathbf 0)$ falls, the layer becoming less
compressible, the two effects adding in \eqref{eq:Dc1}. The Darken form lies above
$D_c$ by $6$, $11$ and $14\%$: on the square lattice the rates are not of gradient
type, so that $R(\mathbf 0)>1$ and $D_c$ is the Darken value divided by
$R(\mathbf 0)$ (Sec.~\ref{sec:measured}). A measurement at a second coverage, of
the dephasing rate or of $D_c$, would test these predictions.

A measurement of the whole decay would reach the sum rule directly. With $F(\K)$
from diffraction, the area beneath the decay, or the peak value of the line
\eqref{eq:Somega}, gives $\tau_c$ and through the quotient \eqref{eq:Rmeas} the
ratio $\Gamma_{\rm eff}/R(\K)$, with no fitted rate. The fitted model predicts
$R=1.43$ at $X$ and $1.03$ at $M$ at $\theta=0.1$, rising to $4.2$ and $1.9$ at
$\theta=0.3$. Symmetric rates, which produce no fast break-up of pairs, give
$R=1.07$ at $X$ and $1.18$ at $M$ at $\theta=0.1$: the area would also decide how
the bonds enter the jump rates.

\section{Conclusions}
\label{sec:concl}

The memory equation of the coherent ISF has been applied to the Ising lattice gas
under Kawasaki dynamics, with the SSF and the hop rate in closed form and the
sum rule confirmed by exact spectra, and tested on CO/Cu(001). The single
exponential of the first order, built from equilibrium quantities alone, gives
the correlation time within ten per cent only at weak coupling,
$-0.8\lesssim\beta\varepsilon\lesssim0.55$ along the channel, nearly independently
of coverage (Fig.~\ref{fig:map}), and the closed-form $R^{(2)}$ locates this
region without the exact spectrum; only there can a measured dephasing rate be
read through the statics and the sum rule. Beyond it the memory is asymmetric:
attraction builds it through the slow dissolution of clusters, which lengthens
the correlation time up to ninefold along the channel and more on the surface as
condensation approaches, whereas a repulsive layer relaxes nearly as a single
mode at its structure peak, along the channel and on the square lattice alike.
The dephasing rate fitted over a time window is not a property of the layer
alone, since it depends on where the window lies; the area beneath the decay,
or the peak of the quasielastic line, gives through \eqref{eq:Rmeas} the hop
rate and $R$ with no fitted rate, and measurements should report it. Along a
channel the dynamics is of gradient type, which puts the width of the spectrum
and the second order of the hierarchy in closed form and makes the collective
diffusion coefficient exact at every coupling; a finite width at long wavelength
identifies such rates, and since the width fixes the short-time curvature of the
decay, \eqref{eq:excess}, the jump rates are in principle accessible to
experiment. For CO/Cu(001), fitted with the two rates of an isolated molecule
and the coupling, the lattice gas reproduces the dephasing rates measured along
both azimuths and requires a NN repulsion of at least $2$--$3\,k_BT$,
$35$--$50$~meV, whatever the jump rates, through the low compressibility it gives
the layer. The measured dephasing rate is close to the inverse correlation time,
the decay being nearly a single exponential beyond a fast part of small weight,
and the fitted model predicts how the dephasing rate and the memory change with
coverage: at $\theta=0.3$, $R$ reaches $4$ at $X$, the dephasing rate at the
$c(2\times2)$ point $M$ falls by the narrowing at the structure peak, and $D_c$
rises tenfold above its constant value in the ideal layer.

The Ising lattice gas assumes discrete jumps between well-defined sites, which
fails for shallow barriers; NN pairwise interactions,
whereas dipolar and substrate-mediated interactions are long-ranged and
three-body terms are common on metals; and a choice of jump rates, which the
CO/Cu(001) data already constrain. Longer-ranged
pair interactions enter only through $v(\K)$ in \eqref{eq:FKrpa} and change the
numbers but not the structure of the results; many-body terms must be built into
the statics. The model is thus a minimal one, adequate for the sign and size of
the memory in adlayers with deep wells and short-ranged interactions. Two
directions remain open: $R(\mathbf 0)$ on a
two-dimensional lattice, where the dynamics is not of gradient type and the
clusters used here admit no small momentum transfer, and jump rates that include
tunnelling, as for hydrogen and deuterium on metals, which change how the bonds
enter the rate but leave the statics and the general results unchanged.

\appendix
\section{Finite-size convergence of the exact spectra}
\label{app:conv}

This appendix shows that the ring and cluster sizes of Sec.~\ref{sec:results}
are sufficient. On a ring of $N$ sites the static averages differ from those of
the infinite chain by terms of order $\eta^{N/2}$, the correlation
\eqref{eq:geometric} wrapping around the ring; for $|\beta\varepsilon|\le2$ the
moments, and with them $\Delta$ and $R^{(2)}$, agree with the closed forms to four
digits from $N=12$. The correlation time, weighted by the slowest modes,
converges more slowly, and its limit is estimated by summing the geometric tail
of its increments with $N$. For repulsion and moderate attraction $R$ is
converged to $0.2$ per cent at $N=12$, and along the contour $R=1.1$ of
Fig.~\ref{fig:map} to half a per cent. For strong attraction the finite ring
underestimates $R$, by $0.9$ per cent at $\theta=0.2$, $\beta\varepsilon=+2$
($N=14$) and by $5.6$ per cent at the strongest corner of the map ($N=12$), so
that these values are lower bounds. On the surface clusters $R$ decreases with
the transverse size, the increments shrinking by a factor of $0.25$ to $0.6$ from
$L_\perp=4$ to $5$; the extrapolated values of Table~\ref{tab:2d} are uncertain by
a few per cent at moderate coupling and by more near condensation, and for
repulsion, where the odd size $L_\perp=5$ frustrates the alternating order, the
spread of the last two sizes is quoted.

\section{The third moment along the channel}
\label{app:m3}

The second order of the hierarchy, Sec.~\ref{sec:modes}, used in closed form in
Secs.~\ref{sec:member} and \ref{sec:memory}, needs the third moment
of the relaxation spectrum besides $m_1=\Omega$ and $m_2$. Along the channel it
follows from the gradient form \eqref{eq:Lgrad}, $\Lc\,\delta n_K=-2(1-\cos x)\,\delta h_K$,
and the self-adjointness of $\Lc$ in the equilibrium scalar product,
\begin{equation}
m_3\;=\;\frac{\avg{(\Lc\delta n_K)^{*}(-\Lc)(\Lc\delta n_K)}}{N\theta F}
\;=\;\frac{4(1-\cos x)^{2}}{\theta F(K)}\;\mathcal D(x),
\qquad
\mathcal D(x)\;=\;\frac1N\avg{\delta h_{K}^{*}(-\Lc)\,\delta h_{K}} ,
\label{eq:m3app}
\end{equation}
with $x$ the phase of Sec.~\ref{sec:lg}. The Dirichlet form $\mathcal D$ is half
the equilibrium sum, over all jumps, of the rate times the squared change of
$\delta h_K$. A jump changes $h$ of \eqref{eq:hfun} only at the four sites around
it, with the rates $w_0$ and $w_1$ of \eqref{eq:w01}, so that $\mathcal D$ is an
average over six consecutive sites, which the Markov chain \eqref{eq:markov}
evaluates as a finite sum over $2^{4}$ configurations. It is a trigonometric
polynomial of degree three in $x$, with coefficients polynomial in
$e^{-\beta\varepsilon}$, $\theta$ and $\eta$. With $m_1$ and $m_2$ of
Sec.~\ref{sec:member}, \eqref{eq:coeffs1} then gives $\Omega_1$ and the
closed-form $R^{(2)}$. Equation~\eqref{eq:m3app} reproduces the third moment of
the exact ring spectra to all digits; the coefficients are available from the
author.

\section*{Author contributions}

S.~Miret-Art\'es: conceptualization, methodology, software, formal analysis,
investigation, writing -- original draft, writing -- review and editing, funding
acquisition.

\section*{Conflicts of interest}

There are no conflicts to declare.

\section*{Data availability}

The data extracted from ref.~\citen{Alexandrowicz2004} and the codes for the
exact spectra and the kinetic Monte Carlo simulations are available from the
author on request.

\section*{Acknowledgements}

This work was supported by the Ministry of Science, Innovation and Universities
(Spain) under Grant No.~PID2023-149406NB-I00.

{\footnotesize
\renewcommand{\refname}{Notes and references}
\makeatletter\renewcommand\@biblabel[1]{#1}\makeatother
}

\end{document}